\documentclass[12pt]{article}

\usepackage{graphicx}
\usepackage{dsfont}
\usepackage{amssymb}
\usepackage{amsmath}
\usepackage{mathrsfs}
\usepackage{bm}

\title{Perfect imaging: they don't do it with mirrors}
\author{Ulf Leonhardt and Sahar Sahebdivan\\
School of Physics and Astronomy, University of St Andrews,\\
North Haugh, St Andrews KY16 9SS, UK
}
\date{\today}
\begin{document}
\maketitle
\begin{abstract}
Imaging with a spherical mirror in empty space is compared with the case when the mirror is filled with the medium of Maxwell's fish eye. Exact time--dependent solutions of Maxwell's equations show that perfect imaging is not achievable with an electrical ideal mirror on its own, but with Maxwell's fish eye in the regime when it implements a curved geometry for full electromagnetic waves.
\end{abstract}


\section{Introduction}

The prospect of perfect imaging with negative refraction \cite{Pendry} has initiated the entire research area of metamaterials that, in turn, inspired the development \cite{LeoConform,PSS} of transformation optics \cite{Shalaev,Review,CCS,LPBook}, the subject of this Special Issue. Yet, ironically, negatively--refracting lenses have never perfectly worked in practice, only ``poor man's lenses'', that are substantially thinner than the wavelength, have shown subwavelength imaging \cite{Fang}. The reason is that negative refraction is only possible in highly dispersive and hence highly dissipative materials \cite{Stockman}; here absorption does not only reduce the intensity but severely limits the resolution of the, theoretically, perfect lens. Alternatives are hyperlenses \cite{Jacob,Liu} that rely on materials with indefinite metric. These are anisotropic materials where one of the eigenvalues of the electric permittivity is negative; these materials thus implement a hyperbolic geometry \cite{Liu} (hence the name hyperlens). Hyperlenses are able to funnel out light from near--fields without losing subwavelength detail, but their resolution is given by their geometric dimensions, and is not unlimited. 

Another example of perfect imaging has been known, as a theoretical idea, since an 1854 paper by Maxwell \cite{Maxwell}. This device, called Maxwell's fish eye, because it reminded Maxwell of the eyes of fish, uses positive refraction. Maxwell's fish eye focuses all light rays emitted from any point in an exact image point; it makes a perfect lens for light rays. Luneburg \cite{Luneburg} discovered that it maps light rays in physical space to rays on a virtual sphere, a curved space; Maxwell's fish eye thus is an early example (and inspiration) of non--Euclidean transformation optics \cite{LeoTyc}. However, it is known since Abbe's theory of imaging that the resolution of optical instrument is limited by the wave nature of light \cite{BornWolf}. Ray optics is not sufficient here, especially in curved geometries, unlike in Euclidean transformation optics \cite{PSS} where geometrical optics \cite{BornWolf} is exact \cite{Review,LPBook}. Analytic solutions of Maxwell's equations for Maxwell's fish eye \cite{Fish,LP} proved theoretically that perfect imaging is possible with positive refraction and recent experiments \cite{Preprints} indicate that it also works in practice.  

Yet perfect imaging with positive refraction \cite{Fish,LP,BMG} challenges \cite{Blaikie,LeoReply,Guenneau,Merlin} some of the accepted wisdom \cite{Pendry} of subwavelength imaging: it does not appear to perform an amplification of evanescent waves \cite{Pendry} and requires a drain at the image in the stationary regime \cite{Fish}. To illuminate some of the perceived paradoxa of perfect imaging, it is instructive to consider the simplest possible case \cite{Merlin}: imaging with a spherical mirror. In this paper, the imaging with such a mirror is compared with the imaging in Maxwell's fish eye \cite{LP}. Exact time--dependent solutions of Maxwell's equations reveal the similarities and characteristic differences between the two cases; the mirror cannot perfectly image, but Maxwell's fish eye can. 

\begin{figure}[h]
\begin{center}
\includegraphics[width=25.0pc]{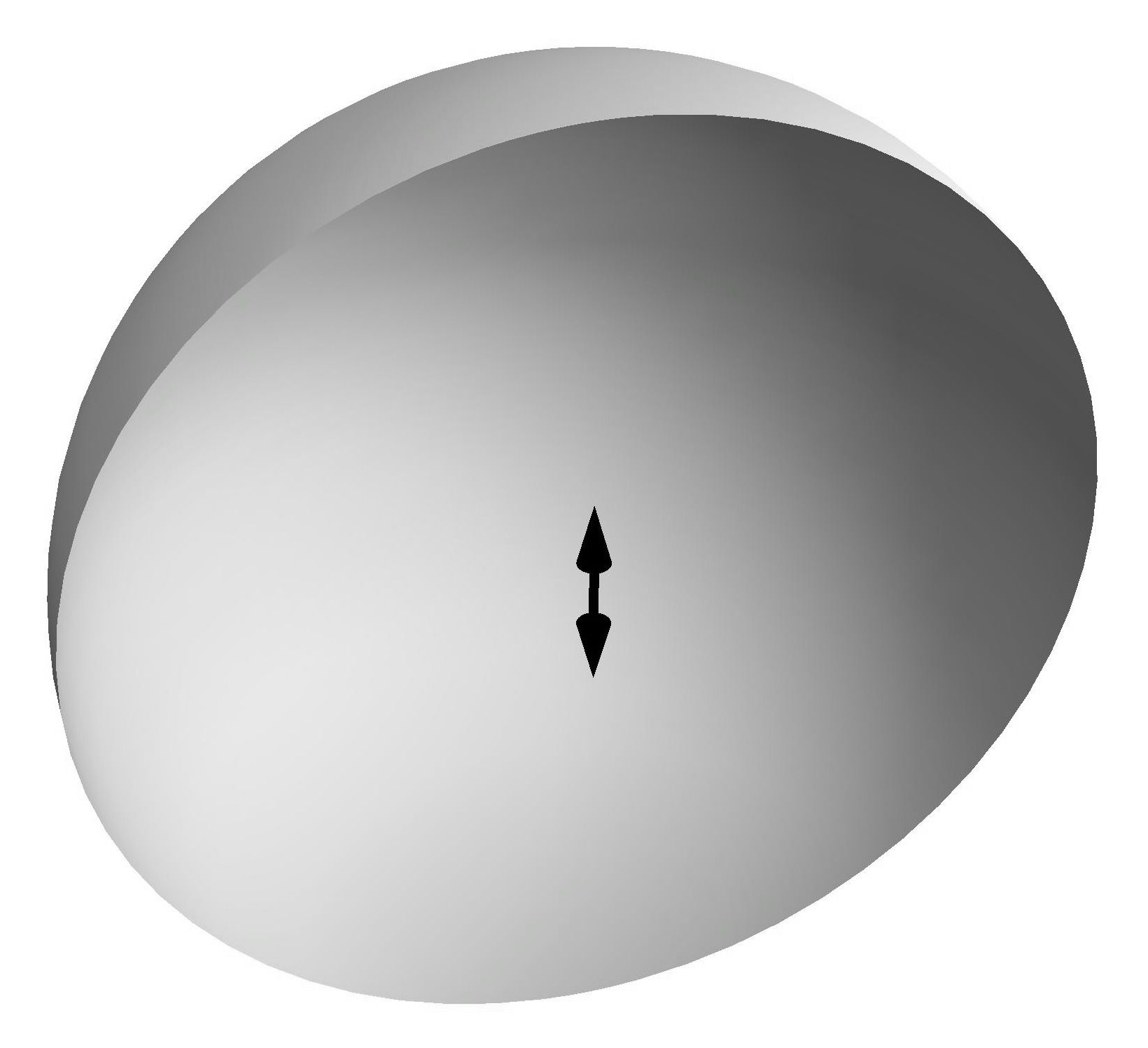}
\caption{
\small{
Spherical mirror. In this paper we consider the electromagnetic wave emitted by a point dipole (double arrow) in the centre of a perfect spherical mirror (half sphere --- only half of the mirror is shown for being able to see the interior). We compare this case with imaging in Maxwell's fish eye in three dimensions \cite{LP}.
}
\label{fig:mirror}}
\end{center}
\end{figure}

Imagine a point dipole placed at the centre of a spherically symmetric mirror that completely surrounds it (Fig.~\ref{fig:mirror}). The dipole emits electromagnetic radiation that reaches the mirror, is reflected there, travels back and focuses at the centre where the dipole sits --- the dipole is imaged to itself. Due to the symmetry of the sphere, this is the simplest conceivable imaging problem. Within the regime of geometrical optics \cite{BornWolf}, the focusing with the mirror would be perfect, but the highly concentrated field in the centre violates the validity condition of geometrical optics \cite{BornWolf}. Therefore, exact calculations of the complete Maxwell equations are required for elucidating the imaging performance of the mirror. The same applies to Maxwell's fish eye \cite{Maxwell} where we fill the space enclosed by the mirror with a dielectric material that has a certain radially symmetric graded refractive index $n(r)$. In both cases, the propagation of electromagnetic waves is governed by a length scale, the size of the spherical mirror, and an associated time scale, the travel time of light in the mirror.  Our results will only depend on these scales. Therefore, to simplify the algebra, it is wise to measure space in units of the mirror radius; in these units the mirror surrounds the dipole at radius
\begin{equation}
\label{mirrorposition}
r = 1 \qquad \text{(mirror).}
\end{equation}
It is also advantageous to measure the time $t$ in units of the distance traveled in free space; in these units we obtain for the speed of light in vacuum
\begin{equation}
\label{c}
c = 1 \,.
\end{equation}
Note that in our units the (angular) frequency $\omega$ is identical to the wavenumber in empty space. We consider the propagation of electromagnetic waves in three--dimensional space, a case where the theory is mathematically simpler than in two--dimensional wave propagation. In 3D, however, impedance matching is required for perfect imaging where the electric permittivity $\varepsilon$ equals the magnetic permeability $\mu$, which is difficult to achieve for materials in practice (but trivially the case in empty space). For the sake of theoretical simplicity we require
\begin{equation}
\label{medium}
\varepsilon = \mu = n(r) \,.
\end{equation}
In empty sphere, the refractive index is unity:
\begin{equation}
\label{empty}
n = 1 \qquad \text{(empty space)}
\end{equation}
and in the case when we fill the sphere with Maxwell's fish eye we have \cite{Maxwell}
\begin{equation}
\label{max}
n = \frac{2}{1+r^2} \qquad \text{(Maxwell's fish eye).}
\end{equation}
Equations (\ref{mirrorposition})-(\ref{max}), together with Maxwell's equations of electromagnetism, set the scene for the problem we investigate in this paper: the radiation of a dipole placed at $r=0$. But before we address this problem a few (well--known) ideas on causality are needed.

\section{Causality}

\begin{figure}[t]
\begin{center}
\includegraphics[width=20.0pc]{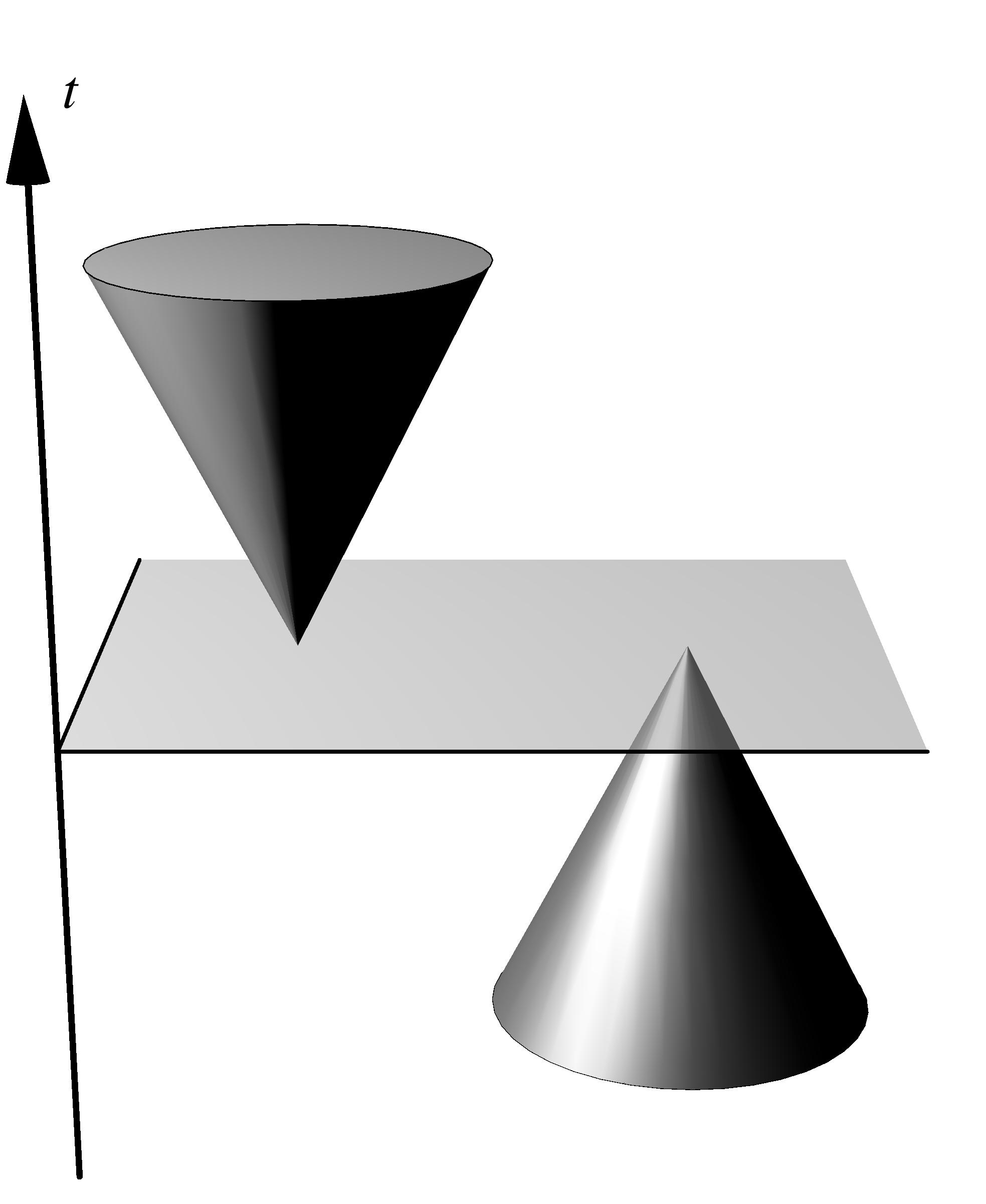}
\caption{
\small{
Schematic space--time diagram of retarded waves (left) and advanced waves (right). The retarded wave is causal; it is created at some moment in time indicated by the planar sheet in the space--time diagram. Prior to that time the wave amplitude is strictly zero. On the other hand, the advanced wave converges into a single spatial point and vanishes there; it is zero after a certain time. Maxwell's equations have both retarded and advanced solutions and superpositions of the two, but only the retarded waves are causal. 
}
\label{fig:causality}}
\end{center}
\end{figure}

Maxwell's electromagnetism is completely time--reversible; a dipole may emit an electromagnetic wave, but it may as well absorb the same wave run in reverse --- by carefully choosing the initial conditions at the boundary an electromagnetic wave may focus at the dipole and be completely absorbed. The latter solutions of Maxwell's equations, where the interaction of the wave with the dipole lies in the future, are called {\it advanced}, whereas the former solutions, where the dipole causes the emission of the wave, are called {\it retarded} (Fig.~\ref{fig:causality}). Clearly, in our case retarded solutions are required; they are causal: no radiation is present prior to the event of emission at, say, time $t=0$. We thus require for all field quantities $G$:
\begin{equation}
\label{causality}
G(t)=0 \quad \mbox{for}\quad t<0 \,.
\end{equation}

\begin{figure}[h]
\begin{center}
\includegraphics[width=20.0pc]{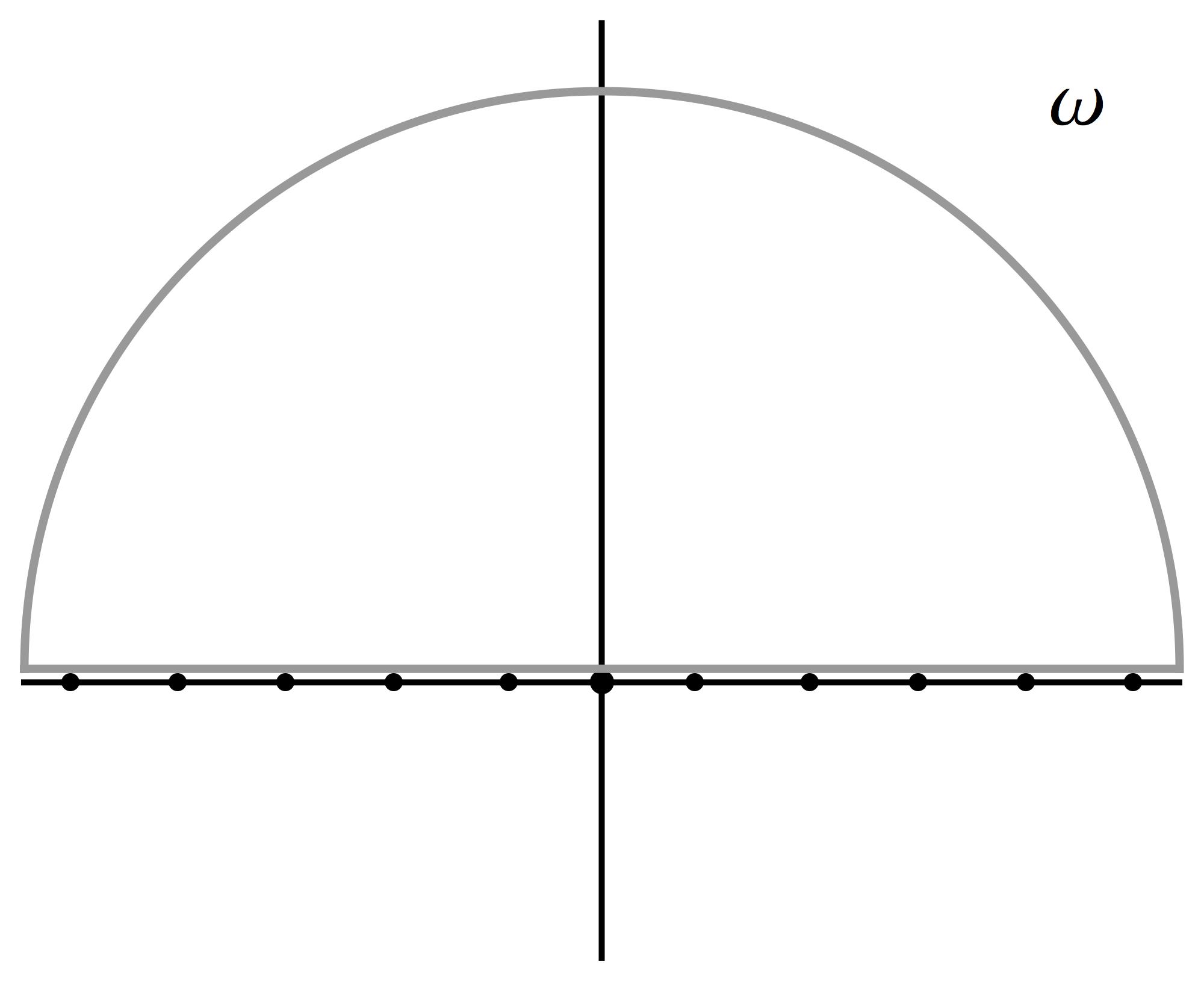}
\caption{
\small{
Causality on the complex frequency plane. The Fourier transform of causal waves (Fig.~\ref{fig:causality}) must be analytic and decaying on the upper half complex $\omega$ plane. The singularities (dots) of the Fourier transform lie on the lower half plane or, as in this picture, on the real axis. In such a case, the contour (grey line) of the Fourier integral (\ref{fourier}) can be closed on the upper half plane for $t<0$ and the integration gives zero, as required by causality (\ref{causality}). The picture shows the first 11 singularities (\ref{poles}) of the Fourier--transformed vector potential of dipole radiation in the spherical mirror (Fig.~\ref{fig:mirror}). 
}
\label{fig:contour}}
\end{center}
\end{figure}

\noindent
Suppose we represent the quantity $G$ in terms of its Fourier components $\widetilde{G}$:
\begin{equation}
\label{fourier}
G(t) = \int_{-\infty}^{+\infty} \widetilde{G}(\omega) \mathrm{e}^{- \mathrm{i} \omega t} \, \mathrm{d}\omega \,.
\end{equation}
The causality condition (\ref{causality}) is met if $\widetilde{G}(\omega)$ is analytic on the upper half complex $\omega$ plane and decays sufficiently fast, such that we can close the integration contour of the Fourier integral (\ref{fourier}) on the upper half plane for $t<0$ when $\exp(-\mathrm{i}\omega t)$ decays for $\mathrm{Im}\,\omega>0$; if $\widetilde{G}(\omega)$ is analytic the closed-contour integral is zero and $G$ vanishes for $t<0$, as required (Fig.~\ref{fig:contour}). Conversely, in the causal case (\ref{causality}) we obtain from the inverse Fourier transformation
\begin{equation}
\label{invfourier}
\widetilde{G}(\omega) = \frac{1}{2\pi} \int_{0}^{+\infty} G(t) \mathrm{e}^{\mathrm{i} \omega t} \, \mathrm{d}t 
\end{equation}
that $\widetilde{G}(\omega)$ obeys the defining Cauchy--Riemann differential equations of analytic functions, because the only frequency--dependent term $\exp(\mathrm{i} \omega t)$ in the integral (\ref{invfourier}) is analytic, provided the integral (\ref{invfourier}) converges. Convergence is guaranteed, because $\mathrm{Re}(\mathrm{i}\omega t)<0$ for $\mathrm{Im}\,\omega>0$ and $t>0$ and so the integrand exponentially decays. Furthermore, in the limit $\mathrm{Im}\,\omega \rightarrow \infty$ the term $\exp(\mathrm{i} \omega t)$ tends to zero, so $\widetilde{G}$ decays at $\infty$ on the upper half plane. We conclude that the Fourier transform of causal fields must be analytic and decaying on the upper half plane. We use this requirement to deduce the causal solutions for our imaging problem.

\section{Symmetry}

Consider a radially--symmetric impedance--matched medium (\ref{medium}) inside the spherical mirror at radius (\ref{mirrorposition}) and suppose that a point dipole is placed at the centre that emits a flash of light. To describe the electromagnetic wave emitted by the dipole we use spherical polar coordinates $\{r,\theta,\phi\}$ with $z$ axis aligned to the dipole. We assume that the vector potential $\bm{A}$ of the wave points in the direction of the dipole with an amplitude $A$ that only depends on the radius $r$ (our calculations will show that this assumption is consistent with Maxwell's equations). The unity vector\footnote{Note that we employ coordinate bases \cite{Review,LPBook} and not the frequently used orthonormal systems of basis vectors that, apart from Cartesian systems, are non--coordinate bases. Coordinate bases are much simpler to work with, given a little knowledge of differential geometry \cite{Review,LPBook}.} $\bm{e}_z$ in $z$--direction reads in terms of the basis vectors in spherical polars \cite{Review,LPBook}
\begin{equation}
\label{ez}
\bm{e}_z = \cos\theta\,\bm{e}_r - \frac{\sin\theta}{r}\,\bm{e_\theta} \,.
\end{equation}
We thus require
\begin{equation}
\label{ansatz0}
A^i = A(r,t) \, \left(\cos\theta, -\frac{\sin\theta}{r}, 0\right) \,.
\end{equation}
However, the vector potential is rather a one--form $A_i$ \cite{LPBook} (also known as a covariant vector) and not the (contravariant) vector $A^i$. We thus lower the index with the metric of the spherical polar coordinates \cite{Review} and write down the ansatz for the vector potential as
\begin{equation}
\label{ansatz}
A_i = A(r,t) \, \left(\cos\theta, -r\sin\theta, 0\right) \,.
\end{equation}
We obtain the magnetic field $\bm{B}$ from $\bm{B}=\bm{\nabla}\times\bm{A}$ where for our ansatz (\ref{ansatz}) we calculate the curl in spherical polars \cite{Review} and lower the index \cite{Review}. We arrive at
\begin{equation}
\label{bs}
B_r = 0\,,\quad B_\theta = 0 \,,\quad B_\phi = B\,\sin^2\theta
\end{equation}
with the magnetic--field amplitude in the plane of emission (for $\theta=\pi/2$) being
\begin{equation}
\label{b}
B = - r \partial_r A \,.
\end{equation}
We use the symbol $\partial_r$ to abbreviate the partial derivative $\partial/\partial r$ and we are going to use  $\partial_t$ for $\partial/\partial t$. The electric field $\bm{E}$ we obtain from Maxwell's equation in the medium (\ref{medium})
\begin{equation}
\label{maxwell}
\bm{\nabla} \times \frac{\bm{B}}{n} - \partial_t\, n \bm{E} = \delta(\bm{r})\delta(t) \,\bm{e}_z
\end{equation}
with our units (\ref{c}) and for a point dipole at the centre pointing in $z$--direction that emits a flash of light at $t=0$ with normalized intensity. We thus find for the non--vanishing electric--field components
\begin{equation}
\label{e}
\partial_t E_r = \left(\frac{2B}{n^2 r^2}-\frac{\delta(\bm{r})\delta(t)}{n}\right) \cos\theta \,,\quad \partial_t  E_\theta = \left(-\frac{1}{n}\,\partial_r\frac{B}{n}+\frac{r \,\delta(\bm{r})\delta(t)}{n}\right) \sin\theta
\end{equation}
while the azimuthal component $\partial_t E_\phi$ is zero. Furthermore, we use Faraday's law $\bm{\nabla}\times \bm{E} = - \partial_t \bm{B}$ expressed in spherical polars \cite{Review} with indices lowered \cite{Review} and our result (\ref{bs}) for the magnetic field, to obtain the wave equation 
\begin{equation}
\label{wave}
\partial_r \frac{1}{n} \,\partial_r \frac{B}{n} - \frac{2B}{n^2r^2} - \partial_t^2 B = r\,\partial_r \frac{\delta(\bm{r})\delta(t)}{n}  \,.
\end{equation}
All other components of $\bm{\nabla}\times \partial_t \bm{E}$ vanish for the electric field (\ref{e}) in spherical polars, which reduces the problem to the wave equation (\ref{wave}) and justifies our ansatz (\ref{ansatz}). Finally, we require that the mirror acts as an ideal electrical mirror where the electric--field components in the mirror are put to zero, in our case (\ref{mirrorposition})
\begin{equation}
\left. E_\theta \right|_{r=1} = 0 \,,
\end{equation}
which implies from expression (\ref{e}) that 
\begin{equation}
\label{mirror}
\left. \partial_r \left(\frac{B}{n}\right) \right|_{r=1} = 0 \,.
\end{equation}
We see that the magnetic field must not change at the mirror. In the following we seek causal solutions of the wave equation (\ref{wave}) with boundary condition (\ref{mirror}) for the magnetic field $B$ expressed in terms (\ref{b}) of the vector potential. First we consider the dipole radiation in empty space surrounded by the spherical mirror \cite{Merlin} and then we compare this case with Maxwell's fish eye \cite{LP}.

\section{Mirror}\label{mirrorsection}

In empty space (\ref{empty}) the left--hand side of the wave equation (\ref{wave}) for the magnetic field (\ref{b}) reduces to 
\begin{equation}
\partial_r^2B-\frac{2B}{r^2}-\partial_t^2 B = -r\,\partial_r\left(\partial_r^2 A+\frac{2}{r}\,\partial_r A-\partial_t^2 A\right)
\end{equation}
such that the wave equation (\ref{wave}) is satisfied if 
\begin{equation}
\label{mirrorwave}
\partial_r^2 A+\frac{2}{r}\,\partial_r A - \partial_t^2 A = -\delta(\bm{r})\delta(t)\,.
\end{equation}
We thus require that the Fourier components $\widetilde{A}$ of the vector potential, with the convention (\ref{fourier}) for the Fourier transform, obey the inhomogeneous wave equation  
\begin{equation}
\label{mirrorwavefourier}
\partial_r^2 \widetilde{A}+\frac{2}{r}\,\partial_r \widetilde{A} + \omega^2 \widetilde{A} = -\frac{\delta(\bm{r})}{2\pi}\,.
\end{equation}
For $r>0$ the general solution of the --- then homogeneous --- wave equation (\ref{mirrorwavefourier}) is
\begin{equation}
\label{sphwaves}
\widetilde{A} = \frac{{\cal A}_+}{r}\,\mathrm{e}^{\mathrm{i}\omega r} + \frac{{\cal A}_-}{r}\,\mathrm{e}^{-\mathrm{i}\omega r} \,.
\end{equation}
Note again that in our units (\ref{c}) the frequency $\omega$ is equal to the wave number that normally appears in the spherical waves (\ref{sphwaves}). The first term of expression (\ref{sphwaves}) describes an outgoing wave and the second term an ingoing wave. The spherical mirror turns outgoing into ingoing waves and so we expect that the coefficients ${\cal A}_\pm$ are not independent. Indeed, we obtain from condition (\ref{mirror}) for the magnetic field (\ref{b}) of the wave (\ref{sphwaves})
\begin{equation}
\label{reflection}
{\cal A}_- = - {\cal A}_+ \mathrm{e}^{2\mathrm{i}\omega} \frac{\omega^2+\mathrm{i}\omega-1}{\omega^2-\mathrm{i}\omega-1} = - {\cal A}_+ \mathrm{e}^{2\mathrm{i}\omega+2\mathrm{i}\delta}
\end{equation}
where $\delta$ denotes the phase
\begin{equation}
\label{delta}
\delta = \arctan\left(\frac{\omega}{\omega^2-1}\right) \,.
\end{equation}
At the mirror (\ref{mirrorposition}) the outgoing component ${\cal A}_+\exp(\mathrm{i}\omega r)$ thus produces the ingoing component ${\cal A}_-\exp(-\mathrm{i}\omega r)$ and vice versa, apart from the extra phase (\ref{delta}). This phase vanishes for high frequencies $\omega\rightarrow\infty$ where the wavelength approaches zero and imaging becomes perfect, regardless whether it is subwavelength--limited or not. For finite $\omega$ the phase (\ref{delta}) is responsible for limiting the resolution of the mirror, as we see next.

So far we considered the general solution of the homogeneous radial wave equation with the boundary condition (\ref{mirror}) at the mirror. Now we turn to the solution that satisfies the inhomogeneous equation (\ref{mirrorwavefourier}) and therefore describes the field generated by the point dipole. For this we write down the following combination of outgoing and ingoing waves (\ref{sphwaves})
\begin{equation}
\label{fouriermirror}
\widetilde{A} = \frac{\sin(\omega-\omega r+\delta)}{8\pi^2r\sin(\omega+\delta)} \,.
\end{equation}
One verifies that this $\widetilde{A}$ obeys the condition (\ref{reflection}) of spherical waves reflected by the mirror. We also see that
\begin{equation}
\label{asymptotics}
\widetilde{A} \sim \frac{1}{8\pi^2r} \quad\text{for}\quad r\rightarrow 0 \,,
\end{equation}
where $(4\pi r)^{-1}$ is the Green function of the Poisson equation,
\begin{equation}
\left(\partial_r^2+\frac{2}{r}\,\partial_r\right) \frac{1}{4\pi r} = \nabla^2 \frac{1}{4\pi r} = -\delta(\bm{r}) \,.
\end{equation}
In the vicinity of the origin we can ignore features of $\widetilde{A}$ on the scale of the wavelength that depend on the frequency $\omega$ and so the inhomogeneous wave equation (\ref{mirrorwavefourier}) reduces to the Poisson equation here. Therefore the wave (\ref{fouriermirror}) with the asymptotics (\ref{asymptotics}) satisfies the wave equation (\ref{mirrorwavefourier}), and we already know that it also obeys the boundary condition (\ref{mirror}). The Fourier components (\ref{fouriermirror}) thus constitute the electromagnetic field of a light flash emitted by a point dipole, provided they are causal, i.e.\ analytic and decaying on the upper half $\omega$ plane. 

Causality is the final point we need to consider for the field (\ref{fouriermirror}). Representing the sine functions in expression (\ref{fouriermirror}) in terms of exponentials we see that $\widetilde{A}\rightarrow 0$ for $\Im\,\omega\rightarrow\infty$. Furthermore, $\widetilde{A}$ is analytic, apart from poles $\omega_m$ on the real axis (Fig.~\ref{fig:contour}) where
\begin{equation}
\label{poles}
\sin(\omega_m + \delta_m) = 0 \,,\quad \delta_m=\delta(\omega_m) \,.
\end{equation}
The Fourier transform $\widetilde{A}$ also has a quadratic singularity at $\omega=0$. In order to obtain a causal solution we must move the poles slightly below the real axis, by providing $\omega$ with an infinitesimal, positive imaginary part. Alternatively, we move the integration contour up from the real axis by an infinitesimal distance (Fig.~\ref{fig:contour}). In both cases, we can close the contour of the Fourier integral (\ref{fourier}) on the upper half plane for $t<0$ and get zero, as required for causal waves (\ref{causality}). For $t>0$ we close the contour on the lower half plane and obtain from Cauchy's theorem
\begin{equation}
\label{mirrorfield}
A = \frac{1}{2\pi^2 r} \sum_{m=1}^\infty \eta_m \sin(\omega_m r)  \sin(\omega_m t) 
\end{equation}
with the coefficients
\begin{equation}
\label{coeff}
\eta_m  = \frac{\omega_m^4-\omega_m^2+1}{\omega_m^4-2\omega_m^2} \,.
\end{equation}
For the light flash emitted by the dipole in the centre of the mirror, we obtained the exact solution (\ref{mirrorfield}) with the pole frequencies (\ref{poles}) and the coefficients (\ref{coeff}). Let us investigate the extent to which this wave images the dipole onto itself.

For perfect imaging we require that an outgoing wave is perfectly converted into an ingoing wave. The wave is focused at the centre and bounces back and forth with a period of twice the time it takes to travel to the mirror, $2$ in our units. As a flash of light is described by a radial delta function, the flash in a perfectly imaging device should be given in terms of the periodic delta function $\Delta(t)$ of period $2$, i.e.\ the kernel of the discrete Fourier transform
\begin{equation}
\label{periodicdelta}
\Delta(t) \equiv \sum_{m=-\infty}^{+\infty} \delta(t-2m)  = \frac{1}{2} \sum_{m=-\infty}^{+\infty} \mathrm{e}^{\mathrm{i}m\pi t} = \frac{1}{2} + \sum_{m=1}^\infty \cos(m\pi t) \,. 
\end{equation}
At the mirror (\ref{mirrorposition}), the wave changes sign and propagates backwards. We thus require for the radial wave $A_0$ in the case of perfect imaging of the dipole onto itself
\begin{equation}
\label{a0}
A_0 = \frac{\Delta(t-r)}{4\pi^2 r} - \frac{\Delta(t+r-2)}{4\pi^2 r} = \frac{\Delta(t-r)}{4\pi^2 r} - \frac{\Delta(t+r)}{4\pi^2 r} \,.
\end{equation}
Writing $A_0$ in terms of the periodic delta function $\Delta$ guarantees that $A_0$ is periodic with period $2$. We obtain from expression (\ref{periodicdelta})
\begin{equation}
\label{a0series}
A_0 = \frac{1}{2\pi^2 r} \sum_{m=1}^\infty \sin(m\pi r)  \sin(m\pi t) \,.
\end{equation}
Formula (\ref{a0series}) strongly resembles our result (\ref{mirrorfield}) for the spherical mirror, but not perfectly. There, instead of the eigenfrequencies $m\pi$ the pole frequencies $\omega_m$ appear and the trigonometric series (\ref{mirrorfield}) contains the coefficients $\eta_m$. The poles (\ref{poles}) depend on the phase (\ref{delta}) that, for large $\omega$, approaches $\omega^{-1}$. In this limit we obtain for the poles (\ref{poles})
\begin{equation}
\omega_m = \pm\left(\frac{m\pi}{2} + \sqrt{\frac{m^2\pi^2}{4}-1}\right) \qquad\text{(integer $m$)}
\end{equation}
that approaches $m\pi$ for $m\rightarrow \infty$. In this limit also $\eta_m\rightarrow 1$, as we see from expression (\ref{coeff}). Consequently, whenever large--$m$ components dominate the series (\ref{a0series}) of the perfect wave, they also dominate the wave (\ref{mirrorfield}) in the spherical mirror. We see from definition (\ref{a0}) that this is the case at the infinitely peaked front of the light flash bouncing back and forth in the mirror. Therefore, the perfectly imaging wave makes the dominant contribution to the electromagnetic wave in the spherical mirror.

\begin{figure}[h]
\begin{center}
\includegraphics[width=30.0pc]{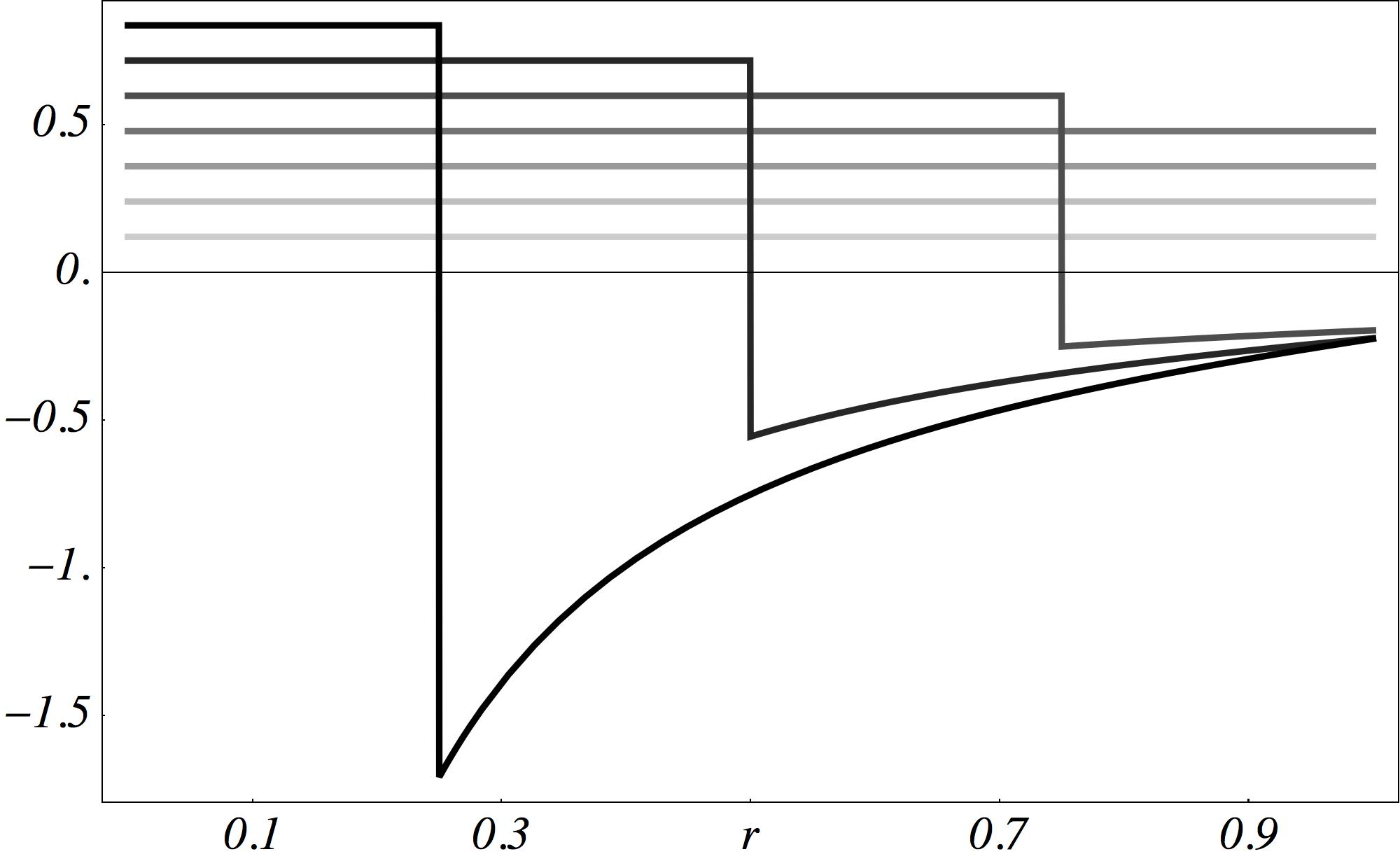}
\caption{
\small{
The reflection of a light pulse in an ideal electrical mirror is not perfect. The figure shows $4\pi(A-A_0)$ where $A$ is the amplitude (\ref{mirrorfield}) of the vector potential of dipole radiation in the spherical mirror and $A_0$ is the amplitude (\ref{a0}) of a perfectly reflected infinitely peaked pulse. The different shades of grey indicate the different evolution times $t=0.25,0.5,0.75,1.0,1.25,1.5,1.75$ in our units, the lighter the tone the earlier is the time. One sees that before the pulse has reached the mirror at $r=1$ the amplitude of the radial vector potential is uniform in space but grows linearly in time. After the interaction with the mirror the amplitude develops an additional discontinuity at the position of the pulse and a field in its wake (see also Fig.~\ref{fig:comparison}).
}
\label{fig:reflection}}
\end{center}
\end{figure}

Figure \ref{fig:reflection} shows the difference $A-A_0$ between the vector potential (\ref{mirrorfield}) of the light reflected in the spherical mirror and perfectly reflected light flashes (\ref{a0}). Relativistic causality implies that the outgoing wave $A$ must remain a perfect pulse until it hits the mirror. Therefore the deviation from the ideal vector potential $A_0$ cannot correspond to any physical field. Figure (\ref{fig:reflection}) indicates that, during this stage, $A-A_0$ is constant in space but grows in time. This growth must be linear in order for $A$ to obey the homogeneous wave equation (\ref{mirrorwave}). As the magnetic field (\ref{b}) and the electric field (\ref{e}) depend on the spatial derivative of $A$ in the gauge relevant to our ansatz (\ref{ansatz}), the electromagnetic fields do not deviate from free dipole radiation until the wave hits the mirror. After the first reflection the wave is no longer a perfect pulse: the vector potential carries an additional discontinuity at the position of the pulse and develops a field in its wake. Note that the mirror is not assumed to be dispersive or otherwise imperfect; these features quantify the imperfections of imaging with an ideal spherical mirror.

\section{Perfection}

Now imagine we fill the space enclosed by the mirror with the impedance--matched (\ref{medium}) medium (\ref{max}) of Maxwell's fish eye \cite{Maxwell}. This device implements the geometry of a curved space. In  particular, it creates the illusion that light propagates on the 3--dimensional surface of the 4--dimensional hypersphere \cite{Luneburg}. It turns out \cite{LP} that the geometry of waves in Maxwell's fish eye appears in the clearest possible form if we write the vector potential $A$ as
\begin{equation}
\label{ad}
A = 2 \int (\partial_r D) \,n \,\mathrm{d}r \quad\text{or, equivalently,}\quad \partial_r A = 2n\partial_r D
\end{equation}
or, from definition (\ref{b}),
\begin{equation}
\label{bd}
B = -2nr\partial_r D \,.
\end{equation}
We show next that the function $D$ is the scalar Green function on the hypersphere \cite{LP}. For this we inspect the left--hand side of the wave equation (\ref{wave}) where
\begin{equation}
\partial_r \frac{1}{n} \,\partial_r \frac{B}{n} - \frac{2B}{n^2r^2} - \partial_t^2 B = -2nr\partial_r\left(\frac{1}{n^3r^2}\,\partial_r n r^2 \partial_r D - D - \partial_t^2 D \right) .
\end{equation}
Therefore the wave equation  (\ref{wave}) for the refractive--index profile (\ref{max}) with $n(0)=2$ is satisfied if $D$ obeys
\begin{equation}
\label{fishwave}
\frac{1}{n^3r^2}\,\partial_r n r^2 \partial_r D - D - \partial_t^2 D = -\frac{\delta(\bm{r})\delta(t)}{n^3}\,,
\end{equation}
or, in Cartesian coordinates,
\begin{equation}
\label{fishwavecartesian}
\frac{1}{n^3}\,\nabla\cdot n \nabla D - D - \partial_t^2 D = -\frac{\delta(\bm{r})\delta(t)}{n^3}\,.
\end{equation}
This is the conformally--coupled radial scalar wave equation on the virtual space implemented by Maxwell's fish eye, the surface of the hypersphere \cite{LP}. If $D$ satisfies the boundary condition (\ref{mirror}) at the mirror in addition to the wave equation (\ref{fishwave}) $D$ is the required scalar Green function. 

Let us construct the Green function $D$. For this we Fourier--transform the wave equation (\ref{fishwavecartesian}),
\begin{equation}
\label{fishfourier}
\frac{1}{n^3}\,\nabla\cdot n \nabla \widetilde{D} + (\omega^2-1) \widetilde{D} = -\frac{\delta(\bm{r})}{2\pi n^3}\,,
\end{equation}
and obtain for the Fourier--transformed radial wave equation:
\begin{equation}
\label{fishwavefourier}
\frac{1}{n^3r^2}\,\partial_r n r^2 \partial_r \widetilde{D} + (\omega^2-1) \widetilde{D} = -\frac{\delta(\bm{r})}{2\pi n^3}\,.
\end{equation}
We consider the limit $r\rightarrow 0$ where the far--field terms $(\omega^2-1)\widetilde{D}$ do not influence the near field of $\widetilde{D}$ anymore, such that we can replace the left--hand side of the Fourier--transformed radial wave equation (\ref{fishwavefourier}) by
\begin{equation}
\frac{1}{n^2r^2}\,\partial_r r^2 \partial_r \widetilde{D} = \frac{1}{n^2}\left(\partial_r^2 + \frac{2\partial_r}{r}\right)\widetilde{D} = \frac{\nabla^2\widetilde{D}}{n^2} = -\frac{\delta(\bm{r})}{2\pi n^3} 
\end{equation}
if $\widetilde{D}$ obeys the asymptotics
\begin{equation}
\label{asympd}
\widetilde{D} \sim \frac{1}{(4\pi)^2 r} \,.
\end{equation}
One easily verifies the following expression for the two fundamental solutions of the Fourier--transformed wave equation (\ref{fishwavefourier}) with asymptotics (\ref{asympd}):
\begin{equation}
\label{sole}
\widetilde{D}_\pm = \frac{1}{(4\pi)^2}\left(r+\frac{1}{r}\right) \exp(\pm 2\mathrm{i}\omega\arctan r) \,.
\end{equation}
Let us write down the expression
\begin{equation}
\label{dfishstat}
\widetilde{D} = \left(r+\frac{1}{r}\right) \frac{\sin(2\omega\,\mathrm{arccot}\, r) - \sin(2\omega\arctan r)}{(4\pi)^2\sin(\pi\omega)} \,.
\end{equation}
We see from Euler's formula $\exp(\mathrm{i}x)=\cos x + \mathrm{i}\sin x$ and $2\,\mathrm{arccot}\,r=\pi-2\arctan r$ that expression (\ref{dfishstat}) is a linear combination of the two fundamental solutions (\ref{sole}) and hence a solution of the homogeneous wave equation (\ref{fishwavefourier}) for $\bm{r}\neq\bm{0}$. Additionally, since $2\,\mathrm{arccot}\,r \rightarrow \pi$ for $r\rightarrow\infty$ expression (\ref{dfishstat}) obeys the required asymptotics (\ref{asympd}) for $\bm{r}\rightarrow\bm{0}$. Furthermore, one verifies that $\widetilde{D}$ conforms to the boundary condition (\ref{mirror}) at the mirror,
\begin{equation}
\left. \partial_r r \partial_r \widetilde{D} \right|_{r=1} = 0 \,.
\end{equation}
Formula (\ref{dfishstat}) thus qualifies as the Fourier transform of a Green function. Finally, $\widetilde{D}$ decays for $\Im\omega \rightarrow \pm\infty$ and is analytic, apart from single poles at
\begin{equation}
\omega_m = m \quad\quad\text{(non--zero integer $m$)}
\end{equation}
and a double pole at $\omega=0$. Similar to the case of the empty spherical mirror (Sec.~\ref{mirrorsection}), we move the polses slightly below the real axis or move the contour of the Fourier integral (\ref{fourier}) above the real axis (Fig.~\ref{fig:contour}) in order to obtain the causal Green function $D$.

\begin{figure}[h]
\begin{center}
\includegraphics[width=35.0pc]{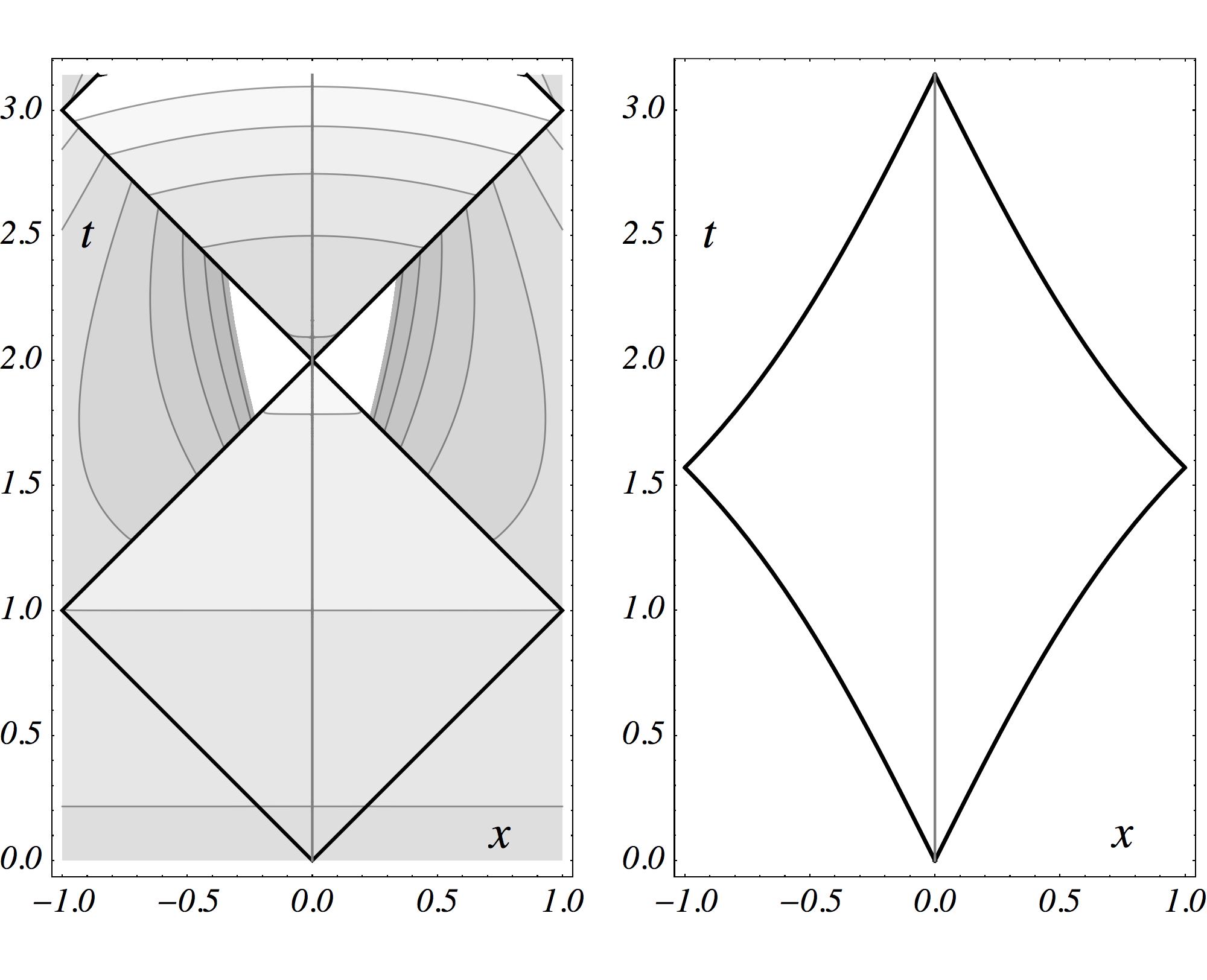}
\caption{
\small{
Mirror versus Maxwell's fish eye. Space--time diagrams of the vector potential of light flashes evolving in time (left) for imaging in a spherical mirror and (right) in Maxwell's fish eye supplemented by a mirror. The mirror is placed at the vertical boundaries of each diagram where the grey vertical line indicates the world line of the centre. The black curves correspond to the infinite peaks of the flashes, the contours describe the finite values of the vector potential. As the right picture contains no finite contours, just infinite peaks, the flash is not disintegrating and bounces back and forth between the mirror in Maxwell's fish eye: imaging is perfect there. In contrast, the left picture shows the imperfections arising from reflection in an ideal electrical mirror: after the flash has bounced off the mirror the vector potential carries an additional discontinuity and a field in its wake that develops a singularity at the refocusing time $t=2$ in our units (the contours are cut off around the singularity). After that time the entire interior of the mirror is filled with a spatially non--uniform vector potential that corresponds to a non--vanishing electromagnetic field distorting the flash: perfect imaging is not done with mirrors.}
\label{fig:comparison}}
\end{center}
\end{figure}

The causal Green function describes the time evolution of a flash of light emitted at the point source in the centre of the fish--eye mirror. We obtain from Cauchy's theorem for the Fourier transform of expression (\ref{dfishstat})
\begin{equation}
D = \frac{\Theta(t)}{4\pi^2}\left(r+\frac{1}{r}\right) \sum_{m=1}^\infty \big[\sin(2m \arctan r) - \sin(2m\,\mathrm{arccot}\,r) \big] \sin (mt)
\end{equation}
and write this result in terms of the periodic delta functions (\ref{periodicdelta}) similar to formula (\ref{a0}) for the ideal imaging wave:
\begin{eqnarray}
D &=& \frac{\Theta(t)}{8\pi^2}\left(r+\frac{1}{r}\right) \left[ \Delta\left(\frac{t-2\arctan r}{\pi}\right)-\Delta\left(\frac{t+2\arctan r}{\pi}\right) \right. \nonumber\\
&& \quad\quad\quad\quad\quad\quad\;\left. - \Delta\left(\frac{t-2\,\mathrm{arccot}\,r}{\pi}\right)+\Delta\left(\frac{t+2\,\mathrm{arccot}\,r}{\pi}\right) \right] \,.
\end{eqnarray}
We write $\Delta$ in terms of the defining series (\ref{periodicdelta}) of delta functions, extract a factor of $\pi$ and combine in the resulting series the numbers $2m$ and $2m+1$ from the $\Delta$ series (\ref{periodicdelta}) in one summation index $m'$. We obtain the result
\begin{equation}
\label{dfishy}
D= \frac{\Theta(t)}{8\pi}\left(r+\frac{1}{r}\right) \sum_{m'=-\infty}^{+\infty} \big[\delta(t-2\arctan r - m'\pi) - \delta(t+2\arctan r - m'\pi)\big] \,.
\end{equation}
Formula (\ref{dfishy}) shows that the light flash emitted by the point dipole propagates through Maxwell's fish eye as a perfect flash. It is reflected at the mirror where the amplitude changes sign, it focuses at the centre and bounces back and forth without disintegrating in time: in contrast to the spherical mirror, Maxwell's fish eye perfectly images in time (Fig.~\ref{fig:comparison}).

\section{Imaging}

We have seen that Maxwell's fish eye faithfully transmits time--dependent signals, it perfectly images in time, but how does it image in space? The spherical mirror focuses only one point, the centre, to itself; an elliptical mirror images one focal point to the other focal point. Maxwell's fish eye is a so--called perfect optical instrument \cite{BornWolf} where all light rays from any source point $\bm{r}_0$ focus at the corresponding image point $\bm{r}_0'$. The fish eye is a perfect instrument for light rays, but the optical resolution of imaging is normally limited by the wave nature of light. Remarkably, one can derive exact solutions of Maxwell's equations for electromagnetic waves in Maxwell's fish eye \cite{LP} (that needs to be impedance--matched in 3D) but the technicalities \cite{LPBook} involved in justifying and explaining them go beyond the scope of this article. The principal technical problem is the tensor nature of the electromagnetic field. Fortunately, we can liberate ourselves from the technicalities of tensor analysis and understand the essence of perfect imaging with positive refraction by discussing the scalar Green function $D$; the electromagnetic field can be derived from $D$ by certain differential forms \cite{LPBook,LP} that reduce to expressions (\ref{bd}) and (\ref{e}) for the special case of the source being at the centre of Maxwell's fish eye. For simplicity, let us also dispense with the mirror and consider the infinitely extended profile (\ref{max}) for $r$ from $0$ to $\infty$. In this case we write down the Fourier--transform of the scalar Green function similar to expression (\ref{dfishstat}), except that we omit the contribution due to the reflection at the mirror,
\begin{equation}
\label{dfish}
\widetilde{D} = \left(r'+\frac{1}{r'}\right) \frac{\sin(2\omega\,\mathrm{arccot}\, r')}{(4\pi)^2\sin(\pi\omega)} \,,
\end{equation}
and replace the radius $r$ by the M\"{o}bius--transformed radius \cite{LPBook,LP}
\begin{equation}
\label{moebius}
r' = \frac{|\bm{r}-\bm{r}_0|}{\sqrt{1+2\bm{r}\cdot\bm{r}_0 + |\bm{r}|^2 |\bm{r}_0|^2}} \,.
\end{equation}
One can show that formulae (\ref{dfish}) and (\ref{moebius}) describe solutions of the Fourier--transformed wave equation (\ref{fishfourier}) with the correct asymptotics at the source point,
\begin{equation}
\widetilde{D} \sim \frac{1}{(4\pi)^2 |\bm{r}-\bm{r}_0|} \,.
\end{equation}
In the ray optics of Maxwell's fish eye \cite{LPBook}, source and image correspond to the $0$ and the $\infty$ of $r'$, the M\"{o}bius--transformed radius (\ref{moebius}):
\begin{eqnarray}
r'= 0 &\leftrightarrow& \bm{r} = \bm{r}_0 \quad\quad\quad\quad\quad\quad\text{(source),} \nonumber\\
r'=\infty &\leftrightarrow& \bm{r} = \bm{r}_0' = -\frac{\bm{r}_0}{|\bm{r}_0|^2} \quad\,\,\,\,\,\text{(image).}
\end{eqnarray}
However, the wave $\widetilde{D}$ is not singular at the image, because we obtain from formula (\ref{dfish}) 
\begin{equation}
\label{spot}
\widetilde{D} \sim \frac{\omega\,\mathrm{sinc}(2\omega/r')}{8\pi^2 \sin(\pi\omega)} \quad \text{with} \quad \mathrm{sinc}\, x = \frac{\sin x}{x} \,.
\end{equation}
We can interpret $\widetilde{D}$ in two ways, as the Fourier transform of the time--dependent Green function $D$ that describes a flash of light or, alternatively, as the amplitude of the continuous wave generated by a stationary point source that emits and absorbs radiation in a stationary state. Our result (\ref{spot}) shows that the electromagnetic wave of a stationary source develops a diffraction--limited image, in agreement with experiment \cite{Preprints}. However, let us adopt the alternative interpretation of $\widetilde{D}$ as the Fourier transform of a light flash $D$. We obtain in complete analogy to the calculations in the previous section:
\begin{equation}
\label{dflashes}
D= \frac{\Theta(t)}{8\pi}\left(r'+\frac{1}{r'}\right) \sum_{m=-\infty}^{+\infty} \big[\delta(t-2\arctan r' - 2m\pi) - \delta(t+2\arctan r' - 2m\pi)\big] \,.
\end{equation}
The flash perfectly focuses at the image point, is reflected there and bounces back to the source etc. In Maxwell's fish eye the image point thus acts like a perfect mirror. In the stationary regime we can imagine the wave as a continuous stream of light flashes. Each of the elementary flashes is reflected at the image point where its amplitude changes sign. In the stationary regime we average over the stream of flashes and so the sign change upon reflection causes the image to get blurred. Maxwell's fish eye has the potential of perfect imaging, but this potential is not realized yet.

What is missing is a crucial ingredient of imaging: a detector. A detector extracts the field at the image point. An ideal point detector acts as a completely passive outlet with point--like resolution (infinitely small cross section). Such a detector could be part of a CCD array or represent the photosensitive molecule of a photographic material that happens to be at the image point. Suppose the detector is positioned at the point $\bm{r}_0$ and extracts the radiation incident at that point. In this case the series (\ref{dflashes}) of flashes bouncing back and forth reduces to one flash that is emitted at the source and disappears at the image at time $t=\pi$ in our units:  
\begin{equation}
\label{dflash}
D= \frac{\Theta(t)}{8\pi}\left(r'+\frac{1}{r'}\right) \delta(t-2\arctan r')\,\Theta(\pi-t) \,.
\end{equation}
For this expression we obtain the Fourier transform
\begin{equation}
\label{drun}
\widetilde{D} = \frac{1}{(4\pi)^2}\left(r'+\frac{1}{r'}\right) \exp(2\mathrm{i}\omega\arctan r') \,.
\end{equation}
Note that formula (\ref{drun}) describes a running wave with complex wave function propagating from the source to the image where the wave disappears, in contrast to the standing wave (\ref{dfish}) with real wave function that is reflected at the image. The spatial singularity of $\widetilde{D}$ for $r'\rightarrow\infty$ corresponds to a supplementary source at the image point $\bm{r}_0'$, a drain. As $2\arctan r'\rightarrow\pi$ for $r'\rightarrow\infty$ the wave carries a phase delay of $\pi\omega$ at the image \cite{LP}. The detector, acting as a drain, creates a perfect spatial image of the source point. So far, we assumed that the detector is positioned at the correct place. Now imagine the detector is elsewhere. As the detector has point--like resolution it can only detect a field that is infinitely concentrated at that point. Therefore, an ideal point detector placed at the wrong position will not detect anything; the perfectly focused field will only appear at the perfectly positioned point detector\footnote{In practice \cite{Preprints} the resolution is limited by the finite electromagnetic cross section of detectors.} in agreement with experiments \cite{Preprints}. Of course, the detector may have a non--perfect efficiency such that it does not extract the entire field. In this case, part of the field is reflected, but the transmitted part is perfectly focused; the detected image is infinitely sharp. As any continuous distribution of sources can be thought of as a distribution of source points, any source is perfectly imaged, as long as it is detected by a continuous array of point detectors. Only the detected field is imaged with point--like precision, but detection is the very point of imaging. 

However, one may still argue that the drain at the image is an artifact that perturbs the field, creating an infinitely focused field entirely on its own, the illusion of perfect imaging but not a true image \cite{Blaikie,Guenneau,Merlin}. Let us consider a counter example where the introduction of a drain turns out not to improve imaging in time: the spherical mirror. Returning to the starting point of this article, we consider empty space enclosed by the spherical mirror. We form the following linear combination of the two fundamental solutions (\ref{sphwaves}) with the reflection condition (\ref{reflection}): 
\begin{equation}
\label{acausal}
\widetilde{A} = \frac{1}{8\pi^2r}\left(\mathrm{e}^{\mathrm{i}\omega r} - \mathrm{e}^{2\mathrm{i}\omega+2\mathrm{i}\delta-\mathrm{i}\omega r}\right) \,,\quad  \mathrm{e}^{2\mathrm{i}\delta}  =  \frac{\omega^2+\mathrm{i}\omega-1}{\omega^2-\mathrm{i}\omega-1} \,.
\end{equation}
This expression creates the singularity of the source at the centre and a drain with phase delay $2\omega+2\delta$ in our units for a wave reflected at the mirror with phase shift $2\delta$ back to the source in time $t=2$. However, expression (\ref{acausal}) also develops poles in the complex frequency plane at the zeros of $\omega^2-\mathrm{i}\omega-1$, i.e.\ at
\begin{equation}
\omega = \frac{\mathrm{i}\pm\sqrt{3}}{2} \,,
\end{equation}
on the upper half plane! Therefore, the solution (\ref{acausal}) is not causal, the drain is not causally connected to the source, but rather acts as an independent source of radiation. However, as we have seen, Maxwell's fish eye makes all the difference here; in this case the drain is consistent with causality.

\newpage

\section{Credo}

An ideal spherical mirror cannot perfectly image on its own, the mirror inevitably distorts light pulses (left of Fig.~\ref{fig:comparison}). However, if the space enclosed by the mirror is filled with the medium of the impedance--matched Maxwell fish eye, pulses are no longer distorted and thus the imaging is perfect in time (right of Fig.~\ref{fig:comparison}). We have seen that the introduction of a drain at the image is allowed by causality and leads to perfect imaging in space in the stationary regime (similar to the perfect focusing of ultrasound waves by time reversal \cite{deRosny}). On the other hand, in imperfect imaging the drain would be in conflict with causality. One of the puzzles of perfect imaging with positive refraction seems to be resolved.

\begin{figure}[h]
\begin{center}
\includegraphics[width=32.0pc]{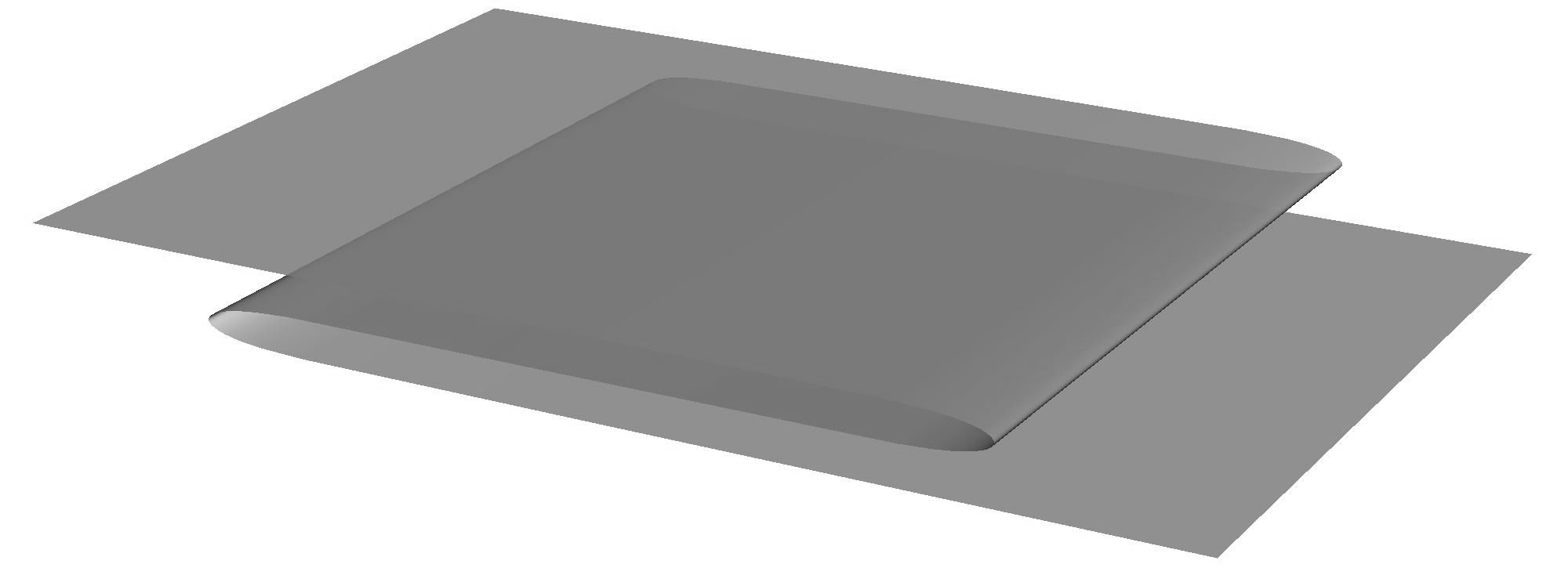}
\caption{
\small{
Negative refraction and folding of space. The figure illustrates the geometrical representation \cite{GREE} of perfect imaging by negative refraction: the device appears to fold space such that the entire electromagnetic field in the folded regions is identical --- the device transfers the field from one region into another: imaging is perfect. Note that this geometrical picture does not discriminate between propagating and evanescent waves.}
\label{fig:folding}}
\end{center}
\end{figure}

Yet one may still wonder how Maxwell's fish eye is able to restore evanescent waves in a similar way negative refraction does \cite{Pendry}. But note that the amplification of evanescent waves is not the only physical picture explaining the performance of the negatively--refracting perfect lens; there is a simple geometrical argument \cite{GREE}: the lens performs a folding transformation of space \cite{CCS}. The negative--index lens appears to fold the space perceived by electromagnetic waves such that three regions are identical, one region in front of the device, the lens itself, and the region behind it (Fig.~\ref{fig:folding}). As these regions are identical for the field, the field strength of an electromagnetic waves must be exactly the same at all triples of connected points: negative refraction makes a perfect lens by spatial transformation \cite{GREE}. In this geometrical picture of imaging we do not need to discriminate between evanescent and propagating waves, because the electromagnetic field is transformed in its entirety. 

\begin{figure}[h]
\begin{center}
\includegraphics[width=32.0pc]{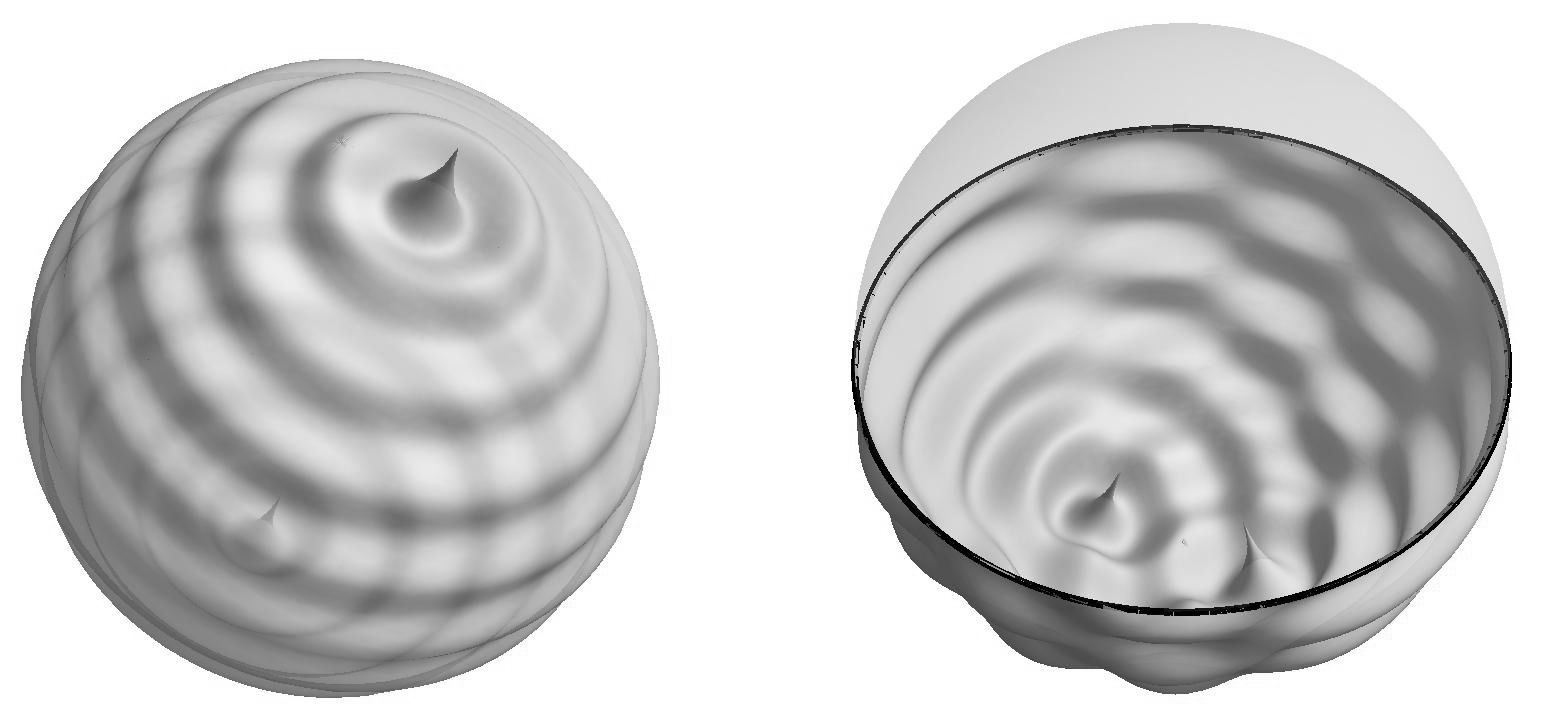}
\caption{
\small{
Geometry behind Maxwell's fish eye. Luneburg \cite{Luneburg} discovered that Maxwell's fish eye implements in 2D the geometry of the two--dimensional surface of a 3D sphere (or in 3D the three--dimensional surface of the 4D hypersphere). The figure illustrates the 2D case \cite{Fish}. The left picture shows a wave on the sphere emitted from a point source and forming a perfect image at its antipodal point. In the right picture a circular mirror is placed at the equator of the sphere such that the image remains on the southern hemisphere \cite{Fish} that corresponds to a finite region in physical space. In 3D a spherical mirror should be used. Source and image merge when the source is placed at the south pole, which in physical space corresponds to emission from the centre of Maxwell's fish eye, the case considered in this paper.}
\label{fig:sphere}}
\end{center}
\end{figure}

Geometry also explains the perfect imaging with Maxwell's fish eye that has a positive refractive--index profile. Like in the case of the perfect lens, this device changes the geometry of space for the electromagnetic field. It creates the effect that electromagnetic waves propagate in the 3--dimensional surface of the 4--dimensional hypersphere. Imagine an ordinary 3--dimensional sphere with 2--dimensional curved surface (the hypersphere is not much different). Suppose light is confined to the surface. There it is completely natural that light waves emitted at any point of the sphere converge on its antipodal point (left of Fig.~\ref{fig:sphere}). In Maxwell's fish eye, the antipodal point on the virtual sphere corresponds to the image point in physical space. In the case discussed in this paper, we cut the fish eye by a spherical mirror that, in virtual space, appears as a mirror around the equator of the sphere (right of Fig.~\ref{fig:sphere}). The introduction of the mirror is not essential for perfect imaging in theory, but it is vital in practice, for limiting the index range and avoiding superluminal propagation \cite{Fish}. The mirror creates the illusion that light explores the entire sphere, whereas in reality it is confined to the semi--sphere where the index range in physical space is smallest. Such conjuring tricks with mirrors are only possible in curved space. Moreover, the imaging of Maxwell's fish--eye, with or without mirror, is not restricted to a single point but happens for all points. This is possible for devices that implement perfect, curved geometries beyond the approximation of geometrical optics. In such cases, the virtual geometry of light created by the material is valid for full electromagnetic waves. Perfect imaging is not done with mirrors \cite{Christie}, but with geometry.

\section*{Acknowledgements}
We thank Roberto Merlin for inspiring this work, Aaron Danner, Susanne Kehr and Tomas Tyc for discussions, and the Royal Society for support.

\end{document}